\documentclass[%
 reprint,
 amsmath,amssymb,
 aps,prl
]{revtex4-2}

\usepackage{graphicx}
\usepackage{dcolumn}
\usepackage{bm}

\begin{document}

\preprint{AAPM/123-QED}

\title{Neutrino Flavor Transformations from New Short-Range Forces}

\author{B.J.P. Jones}
\affiliation{%
University of Texas at Arlington, 108 Science Hall, 502 Yates St, Arlington TX, 76019
}%

\author{J. Spitz}
\affiliation{%
University of Michigan, 450 Church St, Ann Arbor MI 48109
}%

\date{\today}

\begin{abstract}
We examine the commonly explored beyond-standard-model physics scenario of secret neutrino forces, and point out a model prediction that appears to have been
overlooked: the generation of unique flavor-changing effects in experiments featuring decay-at-rest (DAR) neutrino sources. These flavor changes occur because the decay that drives neutrino and antineutrino production, $\mu^{+}\rightarrow e^+ +\bar{\nu}_{\mu}+\nu_{e}$,
is unique in producing two neutrinos in the final state. Any non-flavor-universal
force between the emerging neutrinos would thus induce a new oscillation
phase as they escape from each-other's potential wells, an effect which is largely absent in experiments that primarily rely on meson decay-in-flight and nuclear decay. We calculate the magnitude of the associated observable and compare it to the anomalous neutrino flavor transformation seen by the LSND experiment, finding a wide but constrained allowed parameter space.  We also evaluate existing limits from other experiments, and the testability of this new effect at the future DAR programs JSNS$^2$ and OscSNS.
\end{abstract}

\keywords{Suggested keywords}
\maketitle

\section{Introduction}

The LSND experiment~\cite{Athanassopoulos:1996ds} observed a 3.8$\sigma$ excess of $\bar{\nu}_{e}$-like interactions above the background from a $\pi^+$/$\mu^+$ decay-at-rest (DAR) neutrino source at a distance of 30~m. The excess
of $87.9\pm22.4^{\mathrm{stat}}\pm6.0^{\mathrm{syst}}$ events~\cite{Aguilar:2001ty} can be interpreted in
terms of short-baseline neutrino oscillations caused by
one or more sterile flavors~\cite{abazajian2012light}, with an oscillation probability between
the source and detector of around $0.26\%$.

DAR neutrino experiments, like LSND, take advantage of both a precisely known neutrino ($\nu_\mu,~\nu_e,~\overline{\nu}_\mu$) energy spectrum and interaction cross section from 0-52.8~MeV, and contain minimal contamination ($\sim10^{-3}$) from intrinsic $\bar{\nu}_{e}$~\cite{burman1990monte}.
A DAR source utilizes protons impinging on a target to produce charged pions $\pi^{\pm}$, which quickly come to rest there or in the surrounding shielding to produce neutrinos through the chain: 
\begin{eqnarray}
\pi^{+}&&\rightarrow\mu^{+}+\nu_{\mu}\label{eq:Chain1}\\
\mu^{+}&&\rightarrow e^{+}+\nu_{e}+\bar{\nu}_{\mu}\label{eq:Chain2}
\end{eqnarray}
Notably, no $\bar{\nu}_{e}$ are present in Eqs.~\ref{eq:Chain1} and \ref{eq:Chain2}. 
Therefore, if a significant number of $\bar{\nu}_{e}$, detected via $\bar{\nu}_{e}p\rightarrow e^+ n$, are observed over backgrounds, a natural
hypothesis is that the only available antineutrino $\bar{\nu}_{\mu}$
must have oscillated to $\bar{\nu}_{e}$. The
challenge for this explanation is that the baseline  required to observe a significant number of events from the 4$\pi$ neutrino source, typically 10s of meters, is orders of magnitude too short to produce a discernible oscillation signature, given the known
neutrino mass splittings~\cite{abazajian2012light}. Introducing additional neutrino mass
states with a larger $\Delta m^{2}$ and associated sterile
flavors can produce effects at shorter baselines. But,
severe tensions~\cite{Collin:2016rao,maltoni_michele_2018_1287015} between the allowed regions from positive results
like LSND and MiniBooNE~\cite{Aguilar-Arevalo:2018gpe} and the non-observation of anomalous oscillations in $\nu_{\mu}$
and $\bar{\nu}_{\mu}$ disappearance experiments~\cite{TheIceCube:2016oqi,Adamson:2017uda} mandate the careful consideration
of other explanations, both novel and mundane, for the LSND anomaly.

A beyond-standard-model scenario that has been considered in a variety of contexts invokes ``secret'' interactions~\cite{Ng:2014pca} between new force carriers and
neutrinos, absent for the other standard model particles (recent Refs. include ~\cite{Bakhti:2018avv,Asaadi:2017bhx,Machado:2015sha,Bertuzzo:2017sbj,Huitu:2017vye}).  For example, new forces between active neutrinos can arise in a gauge invariant fashion if the new mediator couples to right handed neutrinos directly, followed by mixing with active neutrinos after electroweak symmetry breaking.   Here, we do not commit to any specific UV completion, but investigate a general consequence of classes of theories with secret neutrino forces for experiments using DAR sources such as LSND.

\section{Flavor Change from New Forces~\label{sec:Transmut}}

DAR-based experiments are unique among oscillation probes in that each $\mu^+$-decay antineutrino is produced alongside a neutrino.  Since these two particles emerge from the decay of a single muon, they are produced in close spatial proximity, separating as they leave the decay vertex. Given new forces between neutrinos, the emerging neutrino and antineutrino will experience a mutual potential
as they separate.  If the couplings of the
neutrinos to the new field are not mass-universal,
the process of escaping from this potential will imbue the neutrino/antineutrino
produced in muon decay with an additional oscillation phase compared
to those produced in, for example, charged-pion decay, where only one neutrino is emitted. In essence, the $\mu^+$ will not
simply decay to $\bar{\nu}_{\mu}$ and $\nu_{e}$, but to slightly flavor-transmuted states.
This transformation occurs within a very short distance of the decay vertex
and is a generic consequence of new, non-universal forces between neutrinos, independent of the detailed properties of the mediator, except coupling strength and mass.

Neutrino flavor oscillations are a consequence of phase differences between propagating
neutrino mass states. This phase difference between states $i$
and $j$ can be expressed as $\phi_{ij}=\int dt\left[E_{i}(t)-E_{j}(t)\right]/\hbar$. 
$i.$ Taylor expanding $E_{i}=\sqrt{p^{2}+m_{i}^{2}}$ in the limit of small $m_{i}^{2}/p^{2}$ yields the standard neutrino oscillation phase:
\begin{equation}
\phi_{ij}=\int dt\frac{\Delta m_{ij}^{2}}{2p}\hbar\approx\frac{\Delta m_{ij}^{2}L}{2E}.
\end{equation}
The latter equality assumes highly relativistic neutrinos,
so $t\sim L$ and $p\sim E$, and adopts natural units
$\hbar=c=1$. Other non-flavor-universal contributions to the neutrino energy will also induce
phase shifts--for example, in sufficiently high density environments neutrinos experience
a matter potential due to the abundance of e$^-$ in matter, as compared to $\mu^-$ and $\tau^-$~\cite{Mikheev:1986gs,Wolfenstein:1977ue}.  The hypothetical role of a phase due to gravitational fields has also
been explored~\cite{Crocker:2003cw,Fornengo:1996ef}. 

\begin{figure}
\begin{centering}
\includegraphics[width=0.9\columnwidth]{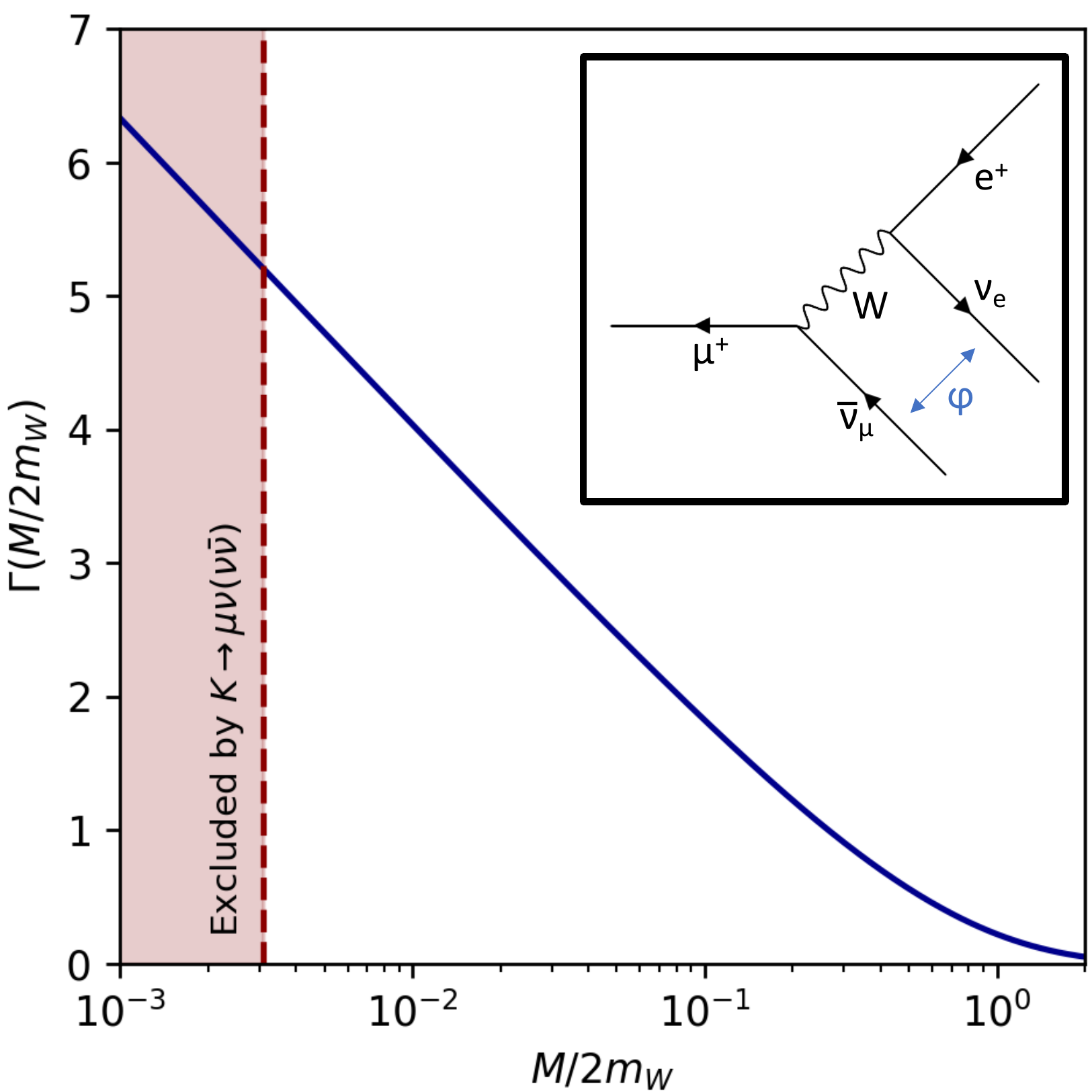}
\par\end{centering}
\vspace{-.3cm}
\caption{The incomplete gamma
function which dictates the scale of the phase shift induced by the new force. Inset: a Feynman diagram from muon decay with the new force indicated.\label{fig:FeynAndGamma}}
\end{figure}

Considering models with new forces between neutrinos, one must also include contributions to the neutrino energy from mutual attraction or repulsion.  The effective
range $r$ of a force carried by a mediator with mass $M$ follows the Yukawa potential~\cite{Yukawa:1935xg}:
\begin{equation}
V_{ab}(r)=\frac{g_{a}g_{b}}{4\pi}\frac{e^{-Mr}}{r},
\end{equation}
where $g_{a}$,~$g_{b}$ are the couplings to each particle.  As discussed below, experimental bounds are strong for mediator masses below $\sim$500~MeV, so the relevant distance scales must be short (1~fm or less). No two neutrinos produced in independent decays are thus expected to reach sufficient proximity to impact each other's oscillations in experimental conditions.  For neutrinos produced in a common decay, however, there is at least momentary proximity where the force will be significant.

To find the contribution of this potential to oscillations, we integrate the quantum phase as the neutrinos escape from each other's potential wells. The total phase is Lorentz invariant, and we  choose to perform the calculation in the reference frame where the recoiling
$\nu_{e}$ is at rest, so the $\nu_{e}$ is considered as a
source of the potential from which the $\bar{\nu}_{\mu}$ escapes.
Because the neutrinos are very light relative to the Q-value of the
decay, we can safely assume that the $\bar{\nu}_{\mu}$ moves at $c$
in this frame at all times. The total phase shift of an antineutrino
mass state $i$ due to a recoiling neutrino mass state $a$ is then:
\begin{equation}
\phi_{ia}=\int_{0}^{\infty}dtV(t)=g_{i}g_{a}\int_{r_{0}}^{\infty}dr\,\frac{e^{-Mr}}{r}.\label{eq:YukawaInt}
\end{equation}
We have assumed the couplings to be non-universal but diagonal in
the mass basis for simplicity of exposition, though non-diagonal couplings can be accommodated with a trivial extension.
In Eq.~\ref{eq:YukawaInt}, $r_{0}$ is the separation at
which the $\nu_e$ and $\bar{\nu}_{\mu}$ are produced. This distance scale is determined by the range of the virtual $W$ boson separating the two vertices in the Feynman diagram of Fig.~\ref{fig:FeynAndGamma}.  Order-of-magnitude estimates suggest the range of the weak force, $r_{0}\sim1/m_{W}$. A more nuanced calculation (Appendix 1) gives $r_{0}\sim1/2m_{W}$ as the mean distance of the Yukawa interaction.  Since our results depend only logarithmically on this scale, the precise pre-factor is somewhat unimportant.

The integral of Eq.~\ref{eq:YukawaInt} can be evaluated to yield:
\begin{equation}
\phi_{ia}=g_{i}g_{a}\Gamma\left(0,\frac{M}{2m_{W}}\right),
\end{equation}
where $\Gamma$ is the incomplete Gamma function, shown in Fig.~\ref{fig:FeynAndGamma}.  To calculate the neutrino transmutation probability we consider the entangled flavor-space
wave functions of the recoiling $\nu_{e}$ and $\bar{\nu}_{\mu}$. The initial state is:
\begin{equation}
|\psi(t_{0})\rangle=|\nu_{e}\rangle\otimes|\bar{\nu}_{\mu}\rangle=\sum_{ia}U_{ea}U^{*}_{\mu i}|\nu_{a}\rangle\otimes|\bar{\nu}_{i}\rangle,
\end{equation}

\begin{figure}
\includegraphics[width=0.99\columnwidth]{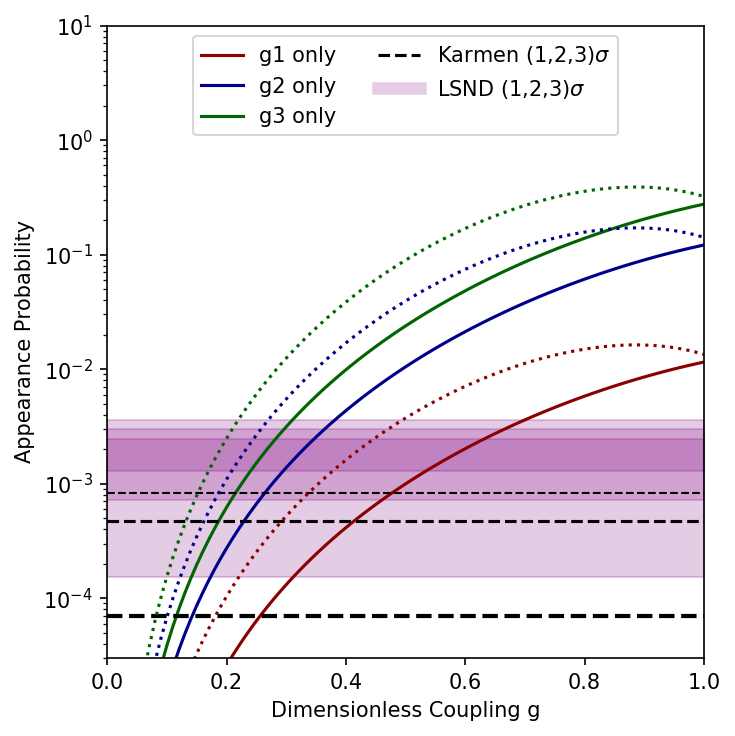}\\
\vspace{-.4cm}
\caption{Appearance probabilities given a single active coupling, with others
set to zero for $\Gamma=2$ (solid lines) and $\Gamma=4$ (dotted lines). \label{fig:Appearance-probabilities-given}}
\end{figure}
with $U$ being the leptonic (PMNS) mixing matrix. Once the antineutrino has effectively escaped the potential
generated by the neutrino in its rest frame, the flavor-space wave function is:
\begin{equation}
=\sum_{ia\alpha\beta}U_{ea}U^{*}_{\mu i}U_{\alpha a}^{*}U_{\beta i}\exp\left[ig_{i}g_{a}\Gamma\left(0,\frac{M}{2m_{W}}\right)\right]|\nu_{\alpha}\rangle\otimes|\bar{\nu}_{\beta}\rangle.
\end{equation}
The flavor change probability for the antineutrino from $\bar{\nu}_{\mu}$
to $\bar{\nu}_{e}$ is given by tracing out the unobserved neutrino:
\begin{eqnarray}
P(\bar{\nu}_{\mu}\rightarrow\bar{\nu}_{e})=\sum_{\gamma}\left|\left(\langle\nu_{\gamma}|\otimes\langle\bar{\nu}_{e}|\right)|\psi(t')\rangle\right|^{2}\quad\quad
\\
=\sum_{\gamma}\left|\sum_{ia}U_{ea}U^{*}_{\mu i}U_{\gamma a}^{*}U_{ei}\exp\left[ig_{i}g_{a}\Gamma\left(0,\frac{M}{2m_{W}}\right)\right]\right|^{2}~\label{eq:OscProb}
\end{eqnarray}
Thus, given $M$, $g_{1}$, $g_{2}$ and $g_{3}$ and the PMNS matrix, we can
calculate the transmutation probability.

As an initial exploration, we consider setting all the $g$'s to zero except for one, and examine the probability to transmute as a function of the remaining
 coupling.   Under this scenario, and given the standard values of the PMNS elements
with normal ordering~\cite{tanabashi2018review}, the appearance probabilities for single non-zero $g$ are shown in Fig.~\ref{fig:Appearance-probabilities-given}. Also shown
is the allowed region band of appearance probabilities consistent with the LSND excess at $\pm1,~2,~3\sigma$, from Ref.~\cite{Aguilar:2001ty}. Natural couplings within the perturbative regime at a scale of $g\sim$0.1 appear apt
to explain the anomalous appearance rate at LSND.  The observed energy spectrum is also consistent: Fig.~\ref{rate} shows the LSND beam data overlaid on the expectation for energy-independent $\overline{\nu}_\mu \rightarrow \overline{\nu}_e$ transmutations, normalized to the LSND excess and accounting for energy resolution~\cite{Athanassopoulos:1996ds}. A simple comparison, neglecting correlations, provides a $\chi ^2=7.8$ for 10 dof
(prob.=0.65).
\begin{figure}
\begin{centering}
\includegraphics[width=0.99\columnwidth]{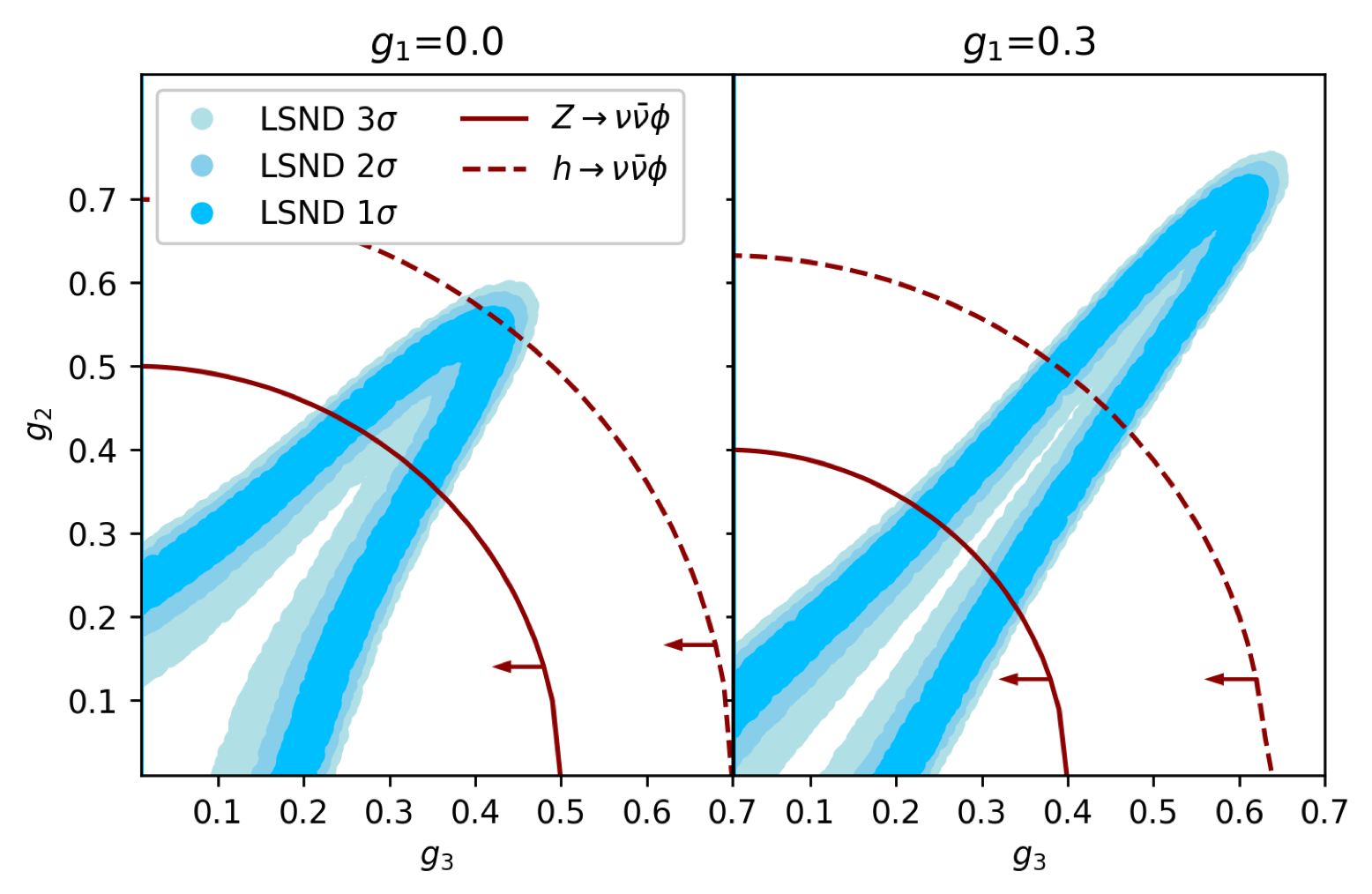}
\par\end{centering}
\vspace{-.4cm}
\caption{LSND allowed regions given a 1~GeV mediator with all $g$'s nonzero.
Each plot explores values of $g_2$ and $g_3$ given a fixed g$_1$ and includes constraints on the invisible Higgs and $Z$ boson widths from Ref.~\cite{Berryman:2018ogk} \label{fig:LSND-allowed-regions}}
\end{figure}

The KARMEN experiment~\cite{Eitel:1998iv} also used a DAR source at  17.7~m to search for short-baseline neutrino oscillations and observed 15 $\overline{\nu}_e$-like events over a predicted background of 15.8 $\pm$ 0.5. This null result presents a constraint on nearly any beyond-standard-model interpretation of LSND, although the KARMEN measurement is statistically much weaker than the LSND observation. Assuming both measurements to have been performed correctly, we can compare them directly.  To estimate the KARMEN constraint, we consider a counting experiment and construct  1, 2, and 3$\sigma$ regions using the Feldman-Cousins approach~\cite{Feldman:1997qc}.   The persistent tension between LSND and KARMEN visible in Fig.~\ref{fig:Appearance-probabilities-given} represents a generic challenge for nearly all explanations of the LSND anomaly, and although the disagreement should not be overlooked, we do not consider it as especially detrimental for the viability of this new physics explanation relative to others.

For a more complete exploration of the parameter space, all three couplings must be allowed to be non-zero. Since $g_{2}$
and $g_{3}$ have the largest impact, we fix $g_{1}$ at discrete values and
make 2D contours in $g_{2}$, $g_{3}$ space. For each point, we
draw a random value of $g_{2}$ and $g_{3}$, calculate the probability of flavor change, and test whether it lies within the allowed LSND region given 1, 2, and 3$\sigma$ statistical fluctuations.
A large and degenerate region of parameter space
appears consistent with LSND (Fig \ref{fig:LSND-allowed-regions}).

While naturalness arguments might favor
order-1 couplings for fundamental interactions, existing experimental limits must be accounted for where applicable.  Constraints on secret neutrino interactions are collected in various works, including Refs.~\cite{Ng:2014pca,Berryman:2018ogk}.  Below the kaon mass, large neutrino-mediator coupling constants are forbidden due to the absence of unobserved processes such as $K^+\rightarrow(\nu\bar{\nu})\mu^+\nu$~\cite{Lessa:2007up}.  These limits effectively rule out scenarios with experimentally observable consequences for $M\leq m_K$, shown as a vertical line in Fig.~\ref{fig:FeynAndGamma}.  For larger mediator masses, experimental constraints are much weaker.  The strongest direct bound for $M\geq 1$~GeV derives from the non-observation of the process $Z\rightarrow \nu {\bar \nu} \phi$ via the invisible width of the $Z$~\cite{zwidth}.  This imposes a requirement that $g<0.5$ given a single-coupling-constant scenario~\cite{Berryman:2018ogk}, which can be extended to multiple couplings by summing over final states (solid line in Fig.~\ref{fig:LSND-allowed-regions}).  A similar but model-dependent constraint, assuming coupling of the new $\phi$ field to the Higgs and the lepton doublets at dimension 6, can also be derived~\cite{Berryman:2018ogk} from the invisible Higgs width (dashed line in Fig.~\ref{fig:LSND-allowed-regions}).  Sub-leading constraints from cosmology, Supernova 1987A, double beta decay, and $D$ meson decays are collected in Ref.~\cite{Berryman:2018ogk}, and found to be much weaker than the invisible $Z$ width constraints in the space of interest.

\section{Exploration with Future Muon Decay-at-Rest Programs}
Neutrino experiments using non-DAR sources are dominated by processes with only one neutrino per decay (in MiniBooNE~\cite{Aguilar-Arevalo:2018gpe}, for example, sub-\% level flux contributions arise from muon decay~\cite{AguilarArevalo:2008yp}), and thus this transmutation effect will be unobservable there. However,
future DAR programs such as OscSNS~\cite{Elnimr:2013wfa} and
JSNS$^{2}$~\cite{Ajimura:2017fld}, both conceived to study the LSND anomaly directly, could observe the effect outlined here.

\begin{table}
\begin{centering}
      \begin{tabular}{|c|c|c|} \hline 
Experiment & OscSNS & JSNS$^2$  \\ \hline \hline \hline
Runtime  &  5 calendar years & 5 calendar years \\ \hline
Baseline  &   60 m & 24 m \\ \hline
POT/year  &  2.2$\times 10^{23}$ & 3.8$\times 10^{22}$ \\ \hline
Proton kinetic energy  & 1 GeV &  3 GeV \\ \hline
$\mu^+$ (DAR) $\overline{\nu}_\mu$/proton  & 0.12 & 0.30  \\ \hline
Fiducial mass  &  450 ton & 17 ton \\ \hline
$\overline{\nu}_e$ efficiency  &  22\% & 38\%  \\ \hline
Intrinsic $\overline{\nu}_e$ fraction  &  $10^{-3}$ & $10^{-3}$ \\ \hline
Intrinsic $\overline{\nu}_e$ norm. unc.  &  30\% & 30\% \\ \hline
\end{tabular} 

\caption{Experimental assumptions used to estimate sensitivity to neutrinophilic forces.\label{experiment_assumptions}}
\end{centering}
\end{table}

\begin{figure}
\begin{centering}
\includegraphics[width=.99\columnwidth]{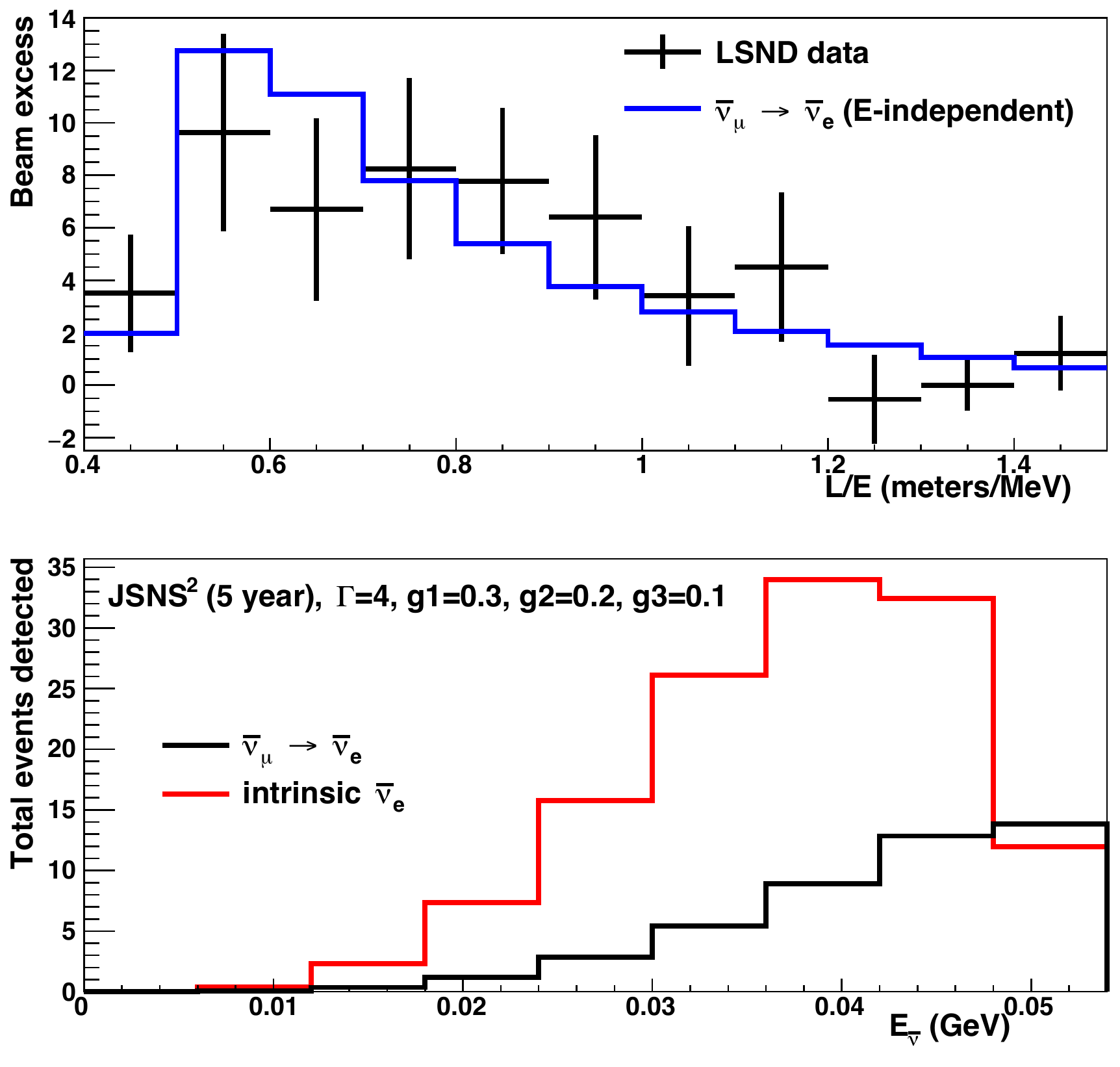}
\par\end{centering}
\vspace{-.4cm}
\caption{(Top) The LSND beam excess overlaid on the distribution expected for energy-independent transmutations. (Bottom) Example transmuted signal and background spectra for JSNS$^2$, given a 5~calendar year exposure and the neutrinophilic force parameters specified. \label{rate}}
\end{figure}

\begin{figure}
\begin{centering}
\includegraphics[width=0.99\columnwidth]{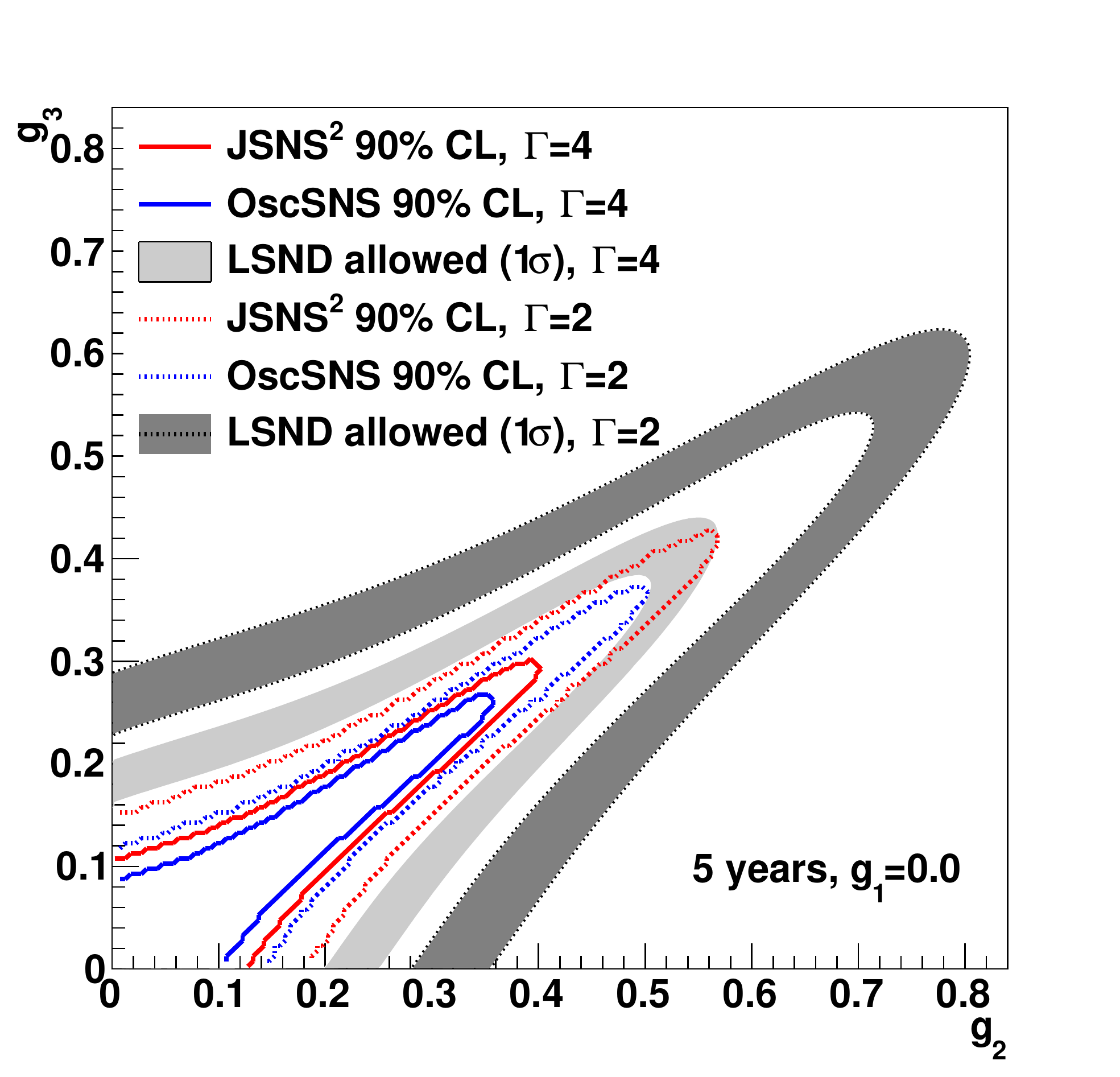}
\par\end{centering}
\vspace{-.4cm}
\caption{The sensitivity of JSNS$^2$ and OscSNS to neutrinophilic forces. \label{sensitivity}}
\end{figure}

We have estimated the sensitivity of both OscSNS and JSNS$^{2}$, given  experimental assumptions shown in Table~\ref{experiment_assumptions}~\cite{Elnimr:2013wfa,Ajimura:2017fld}. The dominant background to a signal of oscillated or transmuted $\overline{\nu}_e$ in both OscSNS and JSNS$^{2}$ is the intrinsic $\overline{\nu}_e$ contamination originating from $\mu^-$ that fails to capture in the target/shielding material, followed by decay ($\mu^- \rightarrow e^- \overline{\nu}_e \nu_\mu$). The background $\overline{\nu}_e$ energy spectrum is fundamentally different than the signal for both energy-independent transmutations and most combinations of possible neutrino oscillation parameters (see Fig.~\ref{rate}), however. Along with the relevant accelerator/source and detector parameters, this intrinsic $\overline{\nu}_e$ fraction and uncertainty ($10^{-3}$ and 30\%, respectively, for both experiments) are crucially important. Fig.~\ref{rate} shows example JSNS$^2$ spectra for the transmuted signal and $\overline{\nu}_e$ background, given a 5~calendar year exposure and an example oscillated signal parameter set ($\Gamma=4$, $g_1=0.3$, $g_2=0.2$, $g_3=0.1$). The sensitivities of JSNS$^2$ and OscSNS, determined using a profile log-likelihood technique~\cite{Rolke:2004mj}  treating the intrinsic $\overline{\nu}_e$ background as a nuisance parameter, are shown in Fig.~\ref{sensitivity} for two example sets of parameters.  The LSND allowed region would be entirely excluded by both programs at more than 90\% CL in both cases.  This conclusion is maintained at all values of $g_1$ and $\Gamma$ explored in the range $0.5<\Gamma<5$ and $0<g_1<0.5$.  Thus, both JSNS$^2$ and OscSNS would provide a definitive statement about this particular interpretation of the LSND anomaly.

\section{Conclusions}

We have re-examined a frequently considered beyond-standard-model scenario in which a neutrino portal connects neutrinos to new secret interactions mediated by massive particles below the weak scale. The implied short-range forces between neutrinos would generate new flavor-changing phases for neutrinos and antineutrinos in close proximity. If this force is non-mass-universal, the emerging neutrino-antineutrino pair in muon decay would produce partially flavor-transformed states.  For mediators in the mass range between the kaon mass and the $W$ mass, \cal{O}(0.1) couplings are weakly constrained experimentally. With couplings of this scale, there is a significant available parameter space that can produce flavor changes consistent with the LSND anomaly.  Tensions with other experimental neutrino programs based on pion or nuclear decay are effectively ameliorated, and future experiments such JSNS$^2$ or OscSNS will be able to sensitively explore this possible effect, distinguishing it from signatures of other beyond-standard-model explanations of the LSND anomaly such as sterile neutrinos.

\section*{Acknowledgements}

We thank Carlos Arg\"uelles, Andr\'e De Gouvêa, Johnathon Jordan, Yoni Kahn, Gordan Krnjaic, Bill Louis, and Pedro Machado for fruitful discussion about this work and their comments on the manuscript.  BJPJ acknowledges the support of the Department of Energy under Early Career Award {DE-SC0019054}. JS is supported by the Department of Energy, Office of Science, under Award No. {DE-SC0007859}.

\bibliographystyle{unsrt}
\bibliography{references}

\section*{Appendix: Initial separation of neutrinos in muon decay}

In this Appendix, we quantitatively motivate that the initial separation of the two neutrinos emerging from muon decay
is \cal{O}$(1/m_{W})$, rather than the alternative distance scale in the problem
\cal{O}$(1/m_{\mu})$.  The amplitudes for particle decays in quantum field theory are traditionally
calculated via the Feynman expansion. The results are scattering amplitudes
from momentum eigenstates in the infinite past to momentum eigenstates
in the infinite future,  erasing all information about positions
and times of interactions in between. This is usually appropriate,
since such intermediate state information is unobservable in, for
example, decay lifetime or scattering cross section calculations. However, it makes the Feynman expansion an imperfect tool for addressing the question at hand.

Feynman diagrams for elastic scattering are often motivated in
textbooks (e.g. Refs.~\cite{halzen88quarks,thomson2013modern}) by comparison to potential scattering with one particle
at rest. It is shown that, for example, the scattering of two particles
via a massless mediator is equivalent to the scattering of one particle
from another as the source of a Coulomb potential. The former approach
has the advantage of making Lorentz invariance explicit while
hiding position-space information, whereas the latter makes the position-space
description explicit while obscuring
Lorentz invariance. Here we will reverse the usual argument, introducing a flavor-changing
potential $V$ for muon decay, by analogy to the case of elastic
scattering. We will ignore spin and its consequences, since we are
only concerned with establishing the distance scale of the propagator.

The flavor-changing potential $V$ includes the creation
and annihilation operators to destroy muons and create the final state
particles. We enforce that these operations
are local, in the sense that particles emerging from a common
vertex must have been annihilated/created at the same position in
space. However, the mediator has a range, encoded in a function $f(\vec{r})$. Thus:
\begin{equation}
V=\int d^{3}\vec{r}d^{3}\vec{y}\,f(\vec{r})\left(a_{\mu}^{\dagger}(\vec{y})a_{\nu}(\vec{y})a_{\bar{\nu}}(\vec{y}-\vec{r})a_{e}(\vec{y}-\vec{r})+h.c.\right)\label{eq:FlavChangingPot}.
\end{equation}
To match the usual Feynman diagram calculation, we would specify in-
and out-going states to be plane waves:
\begin{equation}
|\mu\rangle=\int dx\,e^{ip_{\mu}x_{\mu}}|\mu(x_{\mu})\rangle...
\end{equation}
and so on. Substituting these into Eq. \ref{eq:FlavChangingPot} we can obtain:
\begin{equation}
=\int d^{3}yd^{3}rf(\vec{r})e^{i\left(p_{\mu}y-p_{e}(y-r)-p_{\nu}y-p_{\bar{\nu}}(y-r)\right)}.
\end{equation}
Integrating over $\vec{y}$ gives us a momentum conserving delta function,
leaving only the integral over interaction range $\vec{r}=\vec{x}_{\mu}-\vec{x}_{e}$:
\begin{equation}
=\delta(\vec{p}_{\mu}-\vec{p}_{\nu}-\vec{p}_{e}-\vec{p}_{\bar{\nu}})\int d^{3}\vec{r}f(\vec{r})e^{i\vec{q}\vec{r}},
\end{equation}
where $\vec{q}=\vec{p}_{e}-\vec{p}_{\bar{\nu}}$ is the momentum transfer.
The question then, is what choice of $f(\vec{r})$ will give us the
Feynman propagator for an intermediate particle of mass $m_{W}$?
The answer is the Yukawa form:
\begin{equation}
f=V_{Yukawa}(r)=g_{a}g_{b}\frac{1}{r}e^{-m_{W}r}.
\end{equation}
Which can be easily verified by substitution. Indeed, Yukawa motivated his potential through very
similar arguments regarding the range of the nuclear force. Constructing
the problem in this ``old-fashioned'' way allows us to explicitly
ask questions about spatial properties of the production state.
For example, in this construction we may ask, what is the probability
of producing the electron a distance $\vec{r}'$ away from the muon?
The relevant amplitude would be:
\begin{equation}
\langle\mu(\vec{x}_{\mu})|V|\nu(\vec{x}_{\mu})\bar{\nu}(\vec{x}_{\mu}+\vec{r'})e(\vec{x}_{\mu}+\vec{r}')\rangle=f(\vec{r}'),
\end{equation}
and the probability is the square of this amplitude. To find the mean
range we need to calculate the expectation of the $\hat{r}$ operator,
given a suitably normalized probability distribution, which can be
expressed as:
\begin{equation}
\langle r\rangle=\frac{\int dr\,r|f(r)|^2}{\int dr\,|f(r)|^2}=\frac{2m_{W}}{4m_{W}^{2}}=\frac{1}{2m_{W}}.
\end{equation}
We thus conclude that the expected separation at production is related to the mediator
mass by $\langle r\rangle\sim1/2m_{W}.$
\end{document}